\documentclass[%
aip,
 sd,%
 amsmath,amssymb,
 reprint,%
superscriptaddress]{revtex4-1}
\usepackage{amssymb}
\usepackage{amsmath}
\usepackage{graphicx}
\usepackage{epsfig}
\usepackage{ucs} 
\usepackage{amsmath} 
\usepackage{array}
\usepackage{float}
\usepackage{color}
\usepackage{footmisc}
\usepackage[usenames,dvipsnames]{xcolor}
\usepackage{epstopdf}

\newcommand{\be}{\begin{equation}}
\newcommand{\ee}{\end{equation}}
\newcommand{\bea}{\begin{eqnarray}}
\newcommand{\eea}{\end{eqnarray}}
\newcommand{\beas}{\begin{eqnarray*}}
\newcommand{\eeas}{\end{eqnarray*}}
\newcommand{\nn}{\nonumber}

\begin{document}

\preprint{AIP/123-QED}

\title {In situ photoacoustic characterization for porous silicon growing: detection principles}
\author{C. F. Ramirez-Gutierrez}
\affiliation{Posgrado en Ciencia e Ingenier\'ia de Materiales, Centro de F\'isica Aplicada y Tecnolog\'ia Avanzada, Universidad Nacional Aut\'onoma de M\'exico Campus Juriquilla, C.P. 76230, Quer\'etaro,  Qro., M\'exico}
\affiliation{Licenciatura en Ingenier\'ia F\'isica, Facultad de Ingenier\'ia, Universidad Aut\'onoma de Quer\'etaro, C. P. 76010,Quer\'etaro,  Qro., M\'exico}
\author{J. D. Casta\~no-Yepes}
\affiliation{Instituto de Ciencias Nucleares, Universidad Nacional Aut\'onoma de M\'exico,  M\'exico Distrito Federal, C. P. 04510, M\'exico}
\author{M. E. Rodriguez-Garc\'ia}
\altaffiliation[E-mail corresponding author: ]{marioga@fata.unam.mx}
\affiliation{Licenciatura en Ingenier\'ia F\'isica, Facultad de Ingenier\'ia, Universidad Aut\'onoma de Quer\'etaro, C. P. 76010,Quer\'etaro,  Qro., M\'exico}
\affiliation{Departamento de Nanotecnolog\'ia, Centro de F\'isica Aplicada y Tecnolog\'ia Avanzada, Universidad Nacional Aut\'onoma de M\'exico  Campus Juriquilla, C.P. 76230, Qro., M\'exico}
\date{\today}
\begin{abstract}
There are a few methodologies to monitoring the Porous Silicon (PS) formation in-situ. One of these methodologies is photoacoustic. Previous works that reported the use of photoacoustic to study the PS formation do not provide the physical explanation of the origin of the signal. In this paper, a physical explanation is provided of the origin of the photoacoustic signal during the PS etching.  The incident modulated radiation and changes in the reflectance are taken as thermal sources. In this paper, a useful methodology is proposed to determine the etching rate, porosity, and refractive index of a PS film by the determination of the sample thickness, using SEM images. This method was developed by carrying out two different experiments using the same anodization conditions.  The first experiment consisted of the growth of samples with different etching times to prove the periodicity of the photoacoustic signal and the second considered the growth samples using three different wavelengths that are correlated with the period of the photoacoustic signal. The last experiment showed that the period of the photoacoustic signal is proportional to the laser wavelength.
\end{abstract}

\maketitle

\section{INTRODUCTION}
The electrical and optical properties of porous silicon (PS) have been widely studied in recent years, and it is well know that these properties depend on porosity, and that porosity depends on growing parameters chosen for the electrochemical process. Usually, PS films are characterized after the electrochemical etch using techniques such as scanning electron microscopy (SEM), to obtain surface and cross-section parameters that allow the determination of the thickness and porous size. Profilometry has been used to determine interface roughness, and gravimetry to determine the porosity.\cite{1}  These techniques are contact, destructive, and are not able to give information about physic properties at the same time that the PS  is forming.

One critical aspect of PS in different fields such as optical devices and photonic materials,\cite{2}  gas sensors, \cite{3} biosensors,\cite{4} and composite materials\cite{5} is that the films require specific properties and good optical quality, which imply to control the electrochemical etching in real time. The fabrication of the PS films or devices have  been made by  controlling the electrolyte composition, the current density, and reaction time, but information regarding the growing process (reproducibility) is always neglected. At this point, it is important to note that the reproducibility of the samples with similar properties depends on the operator skills.  For these reasons, it is necessary to develop a methodology to monitor the PS formation in-situ and correlate the electrochemical parameters with its optical properties.

There are a few reports about the in-situ characterization of the PS formation. One of these studies was reported by Foss \textit{et al.}\cite{1} They used an interferometric method by using an infrared laser to illuminate the Si in the side that was not in contact with the electrolyte, producing interference between the beams reflected in the PS/electrolyte, PS/Si, and Si/air interfaces. The obtained signal was analyzed using short-time Fourier transform (STFT), and it was possible to determine PS parameters such as porosity, thickness, etching rate, and interface roughness.  However, the interferometric method cannot provide information about the chemical reaction and changes in the sample temperature, or the formation of multilayer systems.

In contrast, the photoacoustic method is used to study thermal properties of several materials including metals, semiconductors,   insulators, and polymers. In all of the above cases, the thermal properties do not undergo any change as a function of the time. On the other hand, the photoacoustic method has been proven as a suitable and reliable technique to study dynamical processes. The formation of PS represents a dynamical process because the etching produces new interfaces. Therefore, the incident/reflected ratio changes as a function of the optical properties of the PS layers (thickness, porosity, refractive index). This fact implies that the changes in photoacoustic signal are a function of the reflectance changes and changes in thermal properties of the PS-Si multilayer structure. The Photoacoustic signal also provides information about the growing process, as mentioned earlier, at the same time that the electrochemical etching is occurring.

Gutierrez \textit{et al.}\cite{Gutierrez} reported an in-situ technique based on the photoacoustic effect, to characterize the PS formation. However, the physical explanation of the origin and shape of the PA signals during the etching of Si were not described in detail. The data reported by Gutierrez \textit{et al}. and Espinosa-Arbel\'aes \textit{et al.} showed an oscillatory character for the amplitude and phase signals of the photoacoustic signal.\cite{Gutierrez,Espinosa} This shows clearly that the photoacoustic signal depends on various parameters: substrate impurity concentration, current density, and the hydrofluoric acid (HF)/surfactant ratio.

In this study, a physical explanation of the photoacoustic signal during PS formation is provided.  A method to determine   the etching rate, porosity and refractive index of PS films is proposed, helped by SEM images as a complementary technique. The calculations of the changes in the reflectance are based on rules of effective medium for an in-situ characterization of PS formation. Simulation of the changes in the reflectance as a function of the samples thickness and porosity were also studied.

\section{EXPERIMENTAL DETAILS}

\subsection{Electrochemical photoacoustic experimental setup }
To study the kinetics of the porous silicon formation, a differential photoacoustic system (two cells) coupled with an electrochemical cell was developed. The electrochemical system was composed of an electrolyte container, the electrodes, and the current source. The PA amplitude and phase signals were recorded using two Lock-in amplifiers.	

\begin{figure}
\includegraphics[scale=0.3]{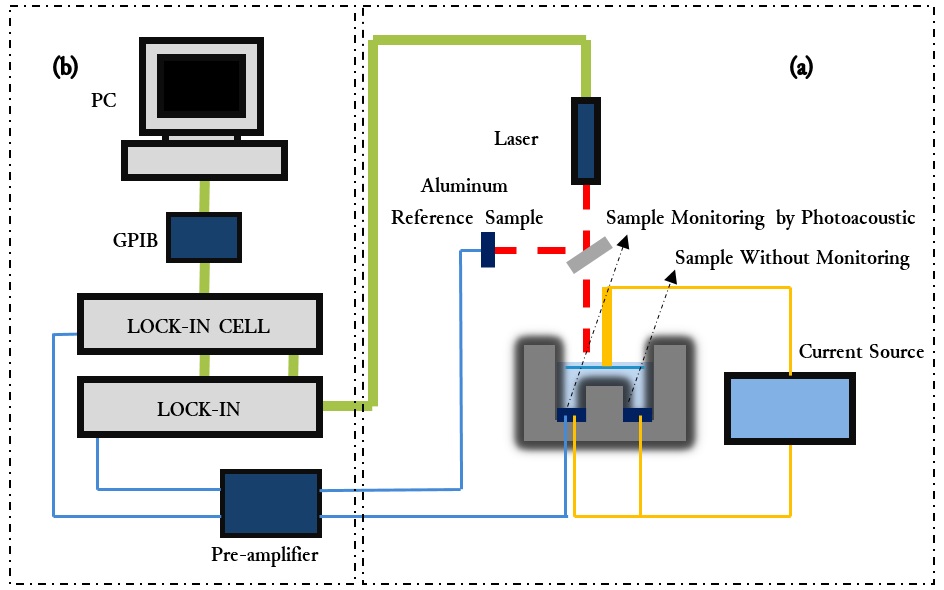}
\caption{Experimental setup used to study the PS growing process. (a) Electrochemical cell coupled to a photoacoustic system  for the etching Si and monitoring \textit{in situ} the etching process respectively. (b) Acquisition data setup.}
\label{Setup}
\end{figure}

Figure~\ref{Setup} shows the experimental setup used to study the PS formation in-situ. Figure~\ref{Setup}(a) shows an electrochemical cell (Polyvinyl chloride, PVC) coupled with a photoacoustic cell. An electrolyte container is located in the upper part of the system. A platinum filament was used as a cathode and a silicon substrate was used as the anode. The cells in the system have the same geometry and dimensions.  A copper foil was in contact with the rear face of the sample. Both electrodes were connected to a controlled current source (6220 Precision Current Source Brand Keithley). The Si substrates were then placed in the photoacoustic cells. With this setup, the growth process of the PS occurred simultaneously in both cells.  A laser (Laser-Mate Group Inc. LBG 8080250 A5-T), 808 nm wavelength, $<200$ mW was used as the excitation source. The excitation beam was modulated at 13 Hz using the lock-in amplifier. A beam splitter with 60/40 was used for the transmitted and reflected beams respectively. The reflected beam was focused on an aluminum film used as a reference sample while the transmitted beam was focused on a silicon sample.  The aluminum sample was used as a means to eliminate any external noise of PA signal.\cite{Espinosa}  In the second cell was another Si sample that grew without PA monitoring, which means that no laser was focused on its surface. Amplitude and phase signals a were recorded with  GPIB acquisition card (Figure \ref{Setup}(b)).

\subsection{Sample description}
Si doped with boron, $<100>$ crystalline orientation, and resistivity $0.001$ $\Omega/\text{cm}^2$ from Polish Corporation of America-USA were used as substrates to growth PS. Substrates were cut into squares of 14 mm in length; each one was washed using the standard method  RCA Silicon wafer cleaning~\cite{RCA1,RCA2}. The current source for the electrochemical etching was fixed at 20 mA. The electrolyte concentration ratio was 3:7 (HF/ethanol in volume), this concentration was selected to provide a good wet surface, allowing homogeneous growth of porous silicon.\cite{Espinosa}

Four samples were etched with the same conditions and with four different etching times. After porous formation, samples were cleaned with ethanol for 20 minutes to remove residues of the electrolyte in the porous structure.

\subsection{Morphological studies}
SEM images were taken to study the morphology of the PS-Si structures, from both cells. A MIRA3 TESCAN microscopy was used, and the analysis was performed using 5.0 kV electron acceleration voltages.  Before the analysis, samples were fixed on a copper specimen holder with carbon tape.  No gold was used to cover the samples prior to the SEM analysis.

\section{PRINCIPLES OF DETECTION}

\subsection{Photoacustic signal generation}

The main mechanism of the photoacoustic signal generation when there are no physicochemical changes in the sample is the absorption of modulating light by the sample. The classical theory of the photoacoustic effect in solids, as described by Rosencwaig and Gershoallows, is to formulate the heat flow equations in the cell resulting from the absorption of energy.\cite{RGPhoto} Figure~\ref{photoCell}(a) shows the geometrical configuration of the PS/Si system. It is characterized by three regions: the region $0<x<l_g$ is the gas chamber, the Si sample  is located in $-(l_b-l_s)<x<0$, and the region $-l_b<x<-(l_b-l_s)$ represents the PS film.  By using this coordinate system, the heat equation in each region can be written as:
\begin{figure}
\includegraphics[scale=0.23]{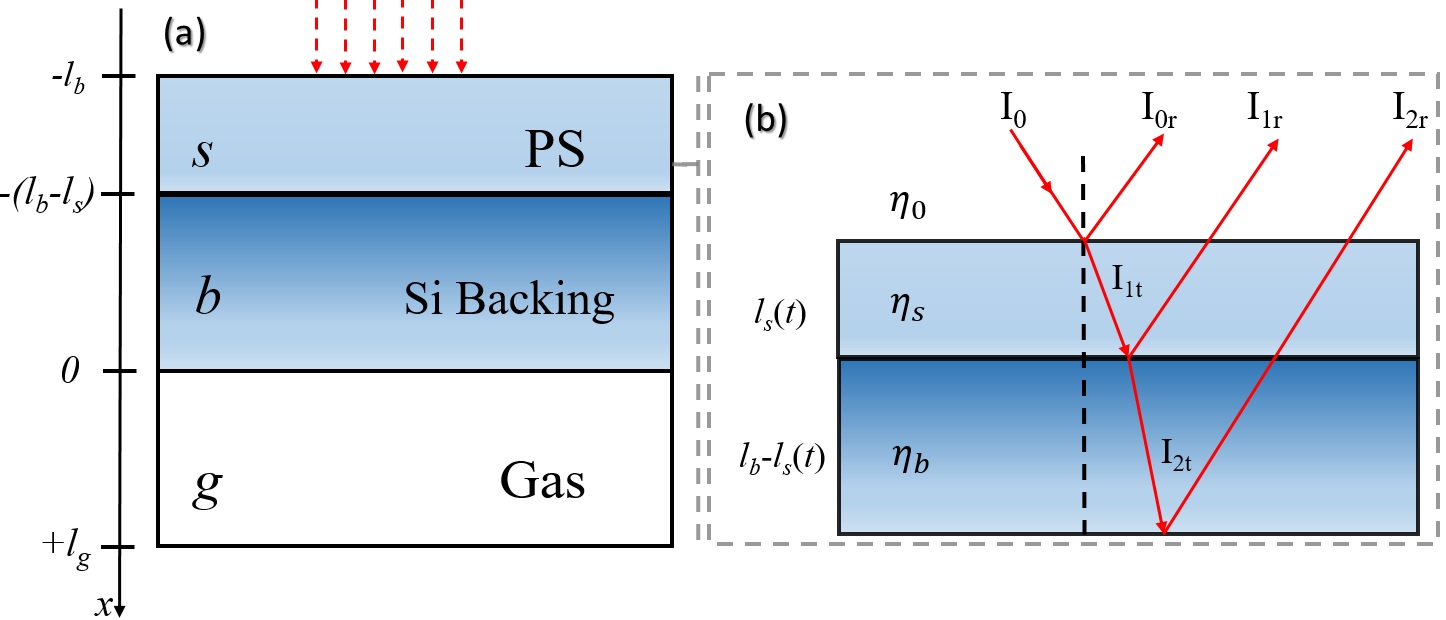}
\caption{ (a) Schematic representation of photoacoustic cell and (b) optical effects.}
\label{photoCell}
\end{figure}
{\small
\bea
\nabla^2T(x,t)-\frac{1}{\alpha}\frac{\partial T(x,t)}{\partial t}=0,\;\; \text{for}\;\; x\leq-l_b,
\label{EDPb}
\eea
{\small
\bea
\nabla^2T_{s,b}(x,t)-\frac{1}{\alpha_{s,b}}\frac{\partial T_{s,b}(x,t)}{\partial t}&=&f(x,t)+Q(t),\nonumber\\
& \text{for}&\;\; -l_b<x\leq 0,
\label{EDPs}
\eea
}
and
\bea 
\nabla^2T_g(x,t)-\frac{1}{\alpha_g}\frac{\partial T_g(x,t)}{\partial t}=0,\;\; \text{for}\;\; 0<x\leq l_g,
\label{EDPg}
\eea
}
where $f(x,t)$  and $Q(t)$ are the heat sources, produced by  the light absorption and electrochemical reaction, respectively. However, the term related with the electrochemical reaction is unknown and is taken as a constant in this study.

The term in relation to the light absorption source has the following form:
\bea
f\left(x,t\right)=\frac{\beta I_0 \eta \left(1-R(t)\right)}{2\kappa_s}e^{\beta x}\left(1+e^{i\omega t}\right),
\label{LSource}
\eea
where $I_o$ is the intensity of the light source, $\omega$ is the modulated frequency, $\beta$  is the absorption coefficient, $\eta$ is the efficiency of the light absorption, $\kappa_s$ is the thermal conductivity, and $R$ the reflectance of the system.

The incidence of the light is by the front of the Si sample, and it is clear that before the etching, $l_s(t=0)=0$. When the source is turned on, the   electrochemical reaction takes place. As a result of the chemical reaction, the thickness of the PS begins to increase, as does the optical path of the light. The interference effects appear to be the product of the creation of the new interface. Thus, the reflectance $(R)$  now has become thickness dependent.
The Figure~\ref{photoCell}(b) shows the schematic representation of the reflection process and the interference effect. Taking into account the above-mentioned effect, the $R$ in the Eq.~\ref{LSource}, is now a function of the etching time.

\subsection{Reflectance Corrections }

The system represented in Figure~\ref{photoCell}(b) is composed of three different mediums where the light can propagated. In the case of the silicon etch, the first medium with a refractive index $\eta_0$ corresponds to the electrolyte composed of Ethanol/HF. The second medium is the PS film with refractive index $\eta_s$, and finally, the third medium corresponds to silicon backing with $\eta_b$.  The interfaces electrolyte/PS and PS/Si-backing can reflect and transmit the laser beam, but the last interface Si-backing/gas can only reflect, and does not permit transmission of radiation for the gas.  Taking into account that the thickness of the Si backing is bigger than the penetration depth of wavelength $\lambda_0$, the fraction of the reflected radiation was determined using a transfer matrix method as follows:\cite{Mitsas}


\begin{figure}
\includegraphics[scale=0.4]{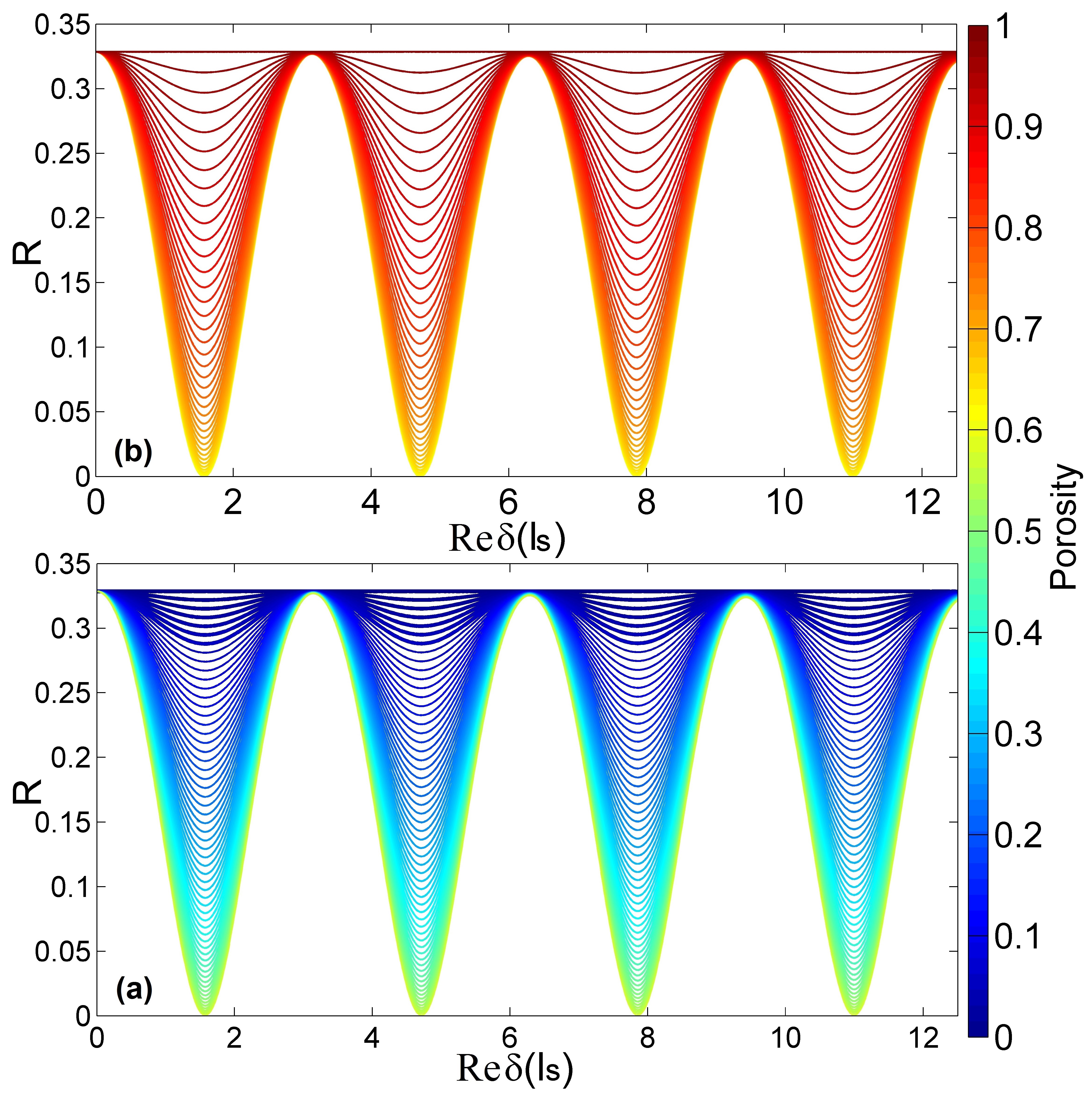}
\caption{Simulation of reflectance as a function of the $Re(\delta)$. The color scale represents the behavior of the reflectance as a function of porosity using a Looyenga~\cite{Looyenga} effective medium  mixture rule for the system electrolyte/PS/Si (Figure \ref{photoCell}(b)). }
\label{Reflectance}
\end{figure}

\bea
R(t)=\left|r^2\right|=\frac{\cos^2(\delta)(\alpha_0-\alpha_1)^2+\sin^2(\delta)(\alpha_0\alpha_1-1)^2}{\cos^2(\delta)(\alpha_0+\alpha_1)^2+\sin^2(\delta)(\alpha_0\alpha_1+1)^2}, \nn\\
\label {RefCoef}
\eea
where $\alpha_0 =\eta_0/\eta_s$ and $\alpha_1 =\eta_b/\eta_s$, using the convention $\eta_x=n_x+i k_x$, where the real part corresponds to the refractive index and the imaginary part  corresponds to the extinction coefficient. 

Here is important to note that $l_s(t)$ represents the change in the sample thickness as a function of the time. However, due to the etching process on the front surface of the substrate, it is necessary to consider and take the  porosity into account. This means that $\eta_s$ is in fact a function of the porosity $p$.

\bea
\delta(t)=\frac{2\pi\eta_s(p) l_s(t)}{\lambda_0},
\label{DELTA}
\eea

Figure~\ref{Reflectance}(a) and~\ref{Reflectance}(b) shows the reflectance changes as a function of the real part of the $\delta (t)$.  The color scale represents the changes in the porosity between 0 to 1. For each one of the simulations, a wavelength of 808 nm was used, and $\eta_s$ was calculated using an effective medium rule for a known porosity.\cite{Looyenga} The product $\eta_s l_s$ that represents the optical path in Eq.~\ref{DELTA}, and these behave as follows: it is possible to have a sample with a high $\eta_s$ and a thin  $l_s$  or a sample with a low $\eta_s$ and thick  $l_s$ and the product is the same. It means that there exist pairs of these parameters that give the same reflectance conditions, for this reason, it is necessary to separate the results into two porosity ranges (see Figure~\ref{Reflectance}).  It is clear that there are values of $\delta(t)$ in which the reflectance changes between maximum and minimum located at   $l_s=\frac{1+2m}{2\eta_s}\lambda_0$  and $l_s=\frac{m}{2\eta_s}\lambda_0 $. These values show that the reflectance changes as a function of the sample thickness and it is modulated by the relationship between the incident wavelength and the change in the thickness of the sample. This fact will be correlated with the photoacoustic signal for the PS formation in the next section.

\section {RESULTS}

This section shows the results obtained from two different experiments: the first experiment consists of the etching of four samples of silicon with the same anodization current and electrolyte composition when the etching time is changed, in order to prove that the photoacoustic signal is self-modulated by the changes in the reflectance of the heterostructure and that it is periodic. The second experiment consisted of the etching the three samples of silicon using a different wavelength of the laser, to prove the relation shift between the wavelength and the period of the photoacoustic signal.

\subsection {Time dependence of photoacoustic signal}

\begin{figure}
\includegraphics[scale=0.35]{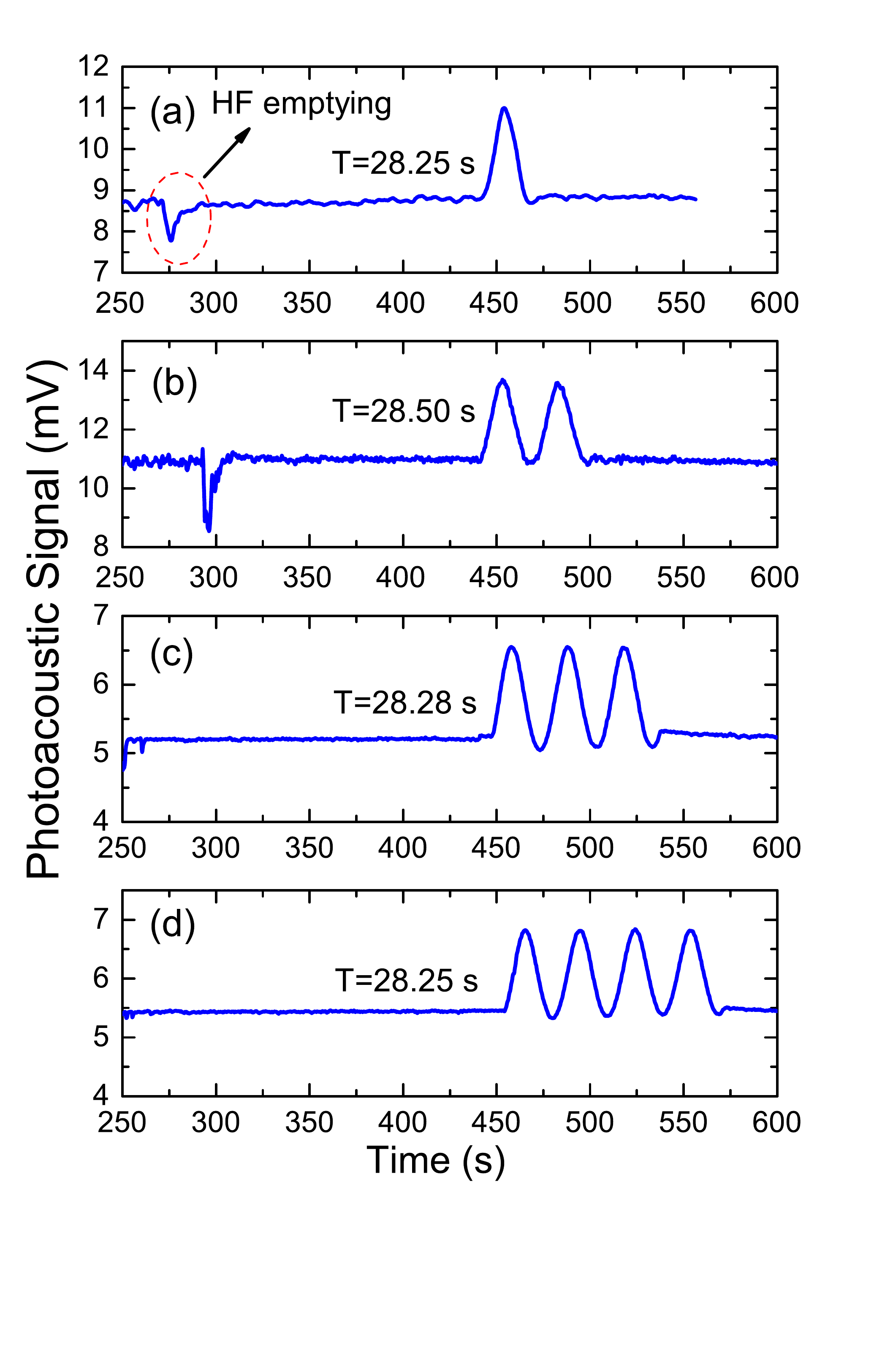}
\caption{Photoacoustic signals for four different etching times at $20$ mA current anodization. Each one of the etching process was stopped at 1(a), 2(b), 3(c), and 4(c) photoacoustic cycles.}
\label{Photoacustic1234}
\end{figure}

Figure~\ref{Photoacustic1234} shows the amplitude of the photoacoustic signal as a function of the time for four different growing processes using the same experimental conditions. In Figure~\ref{Photoacustic1234}(a) the change in the amplitude signal labeled with the red circle represents the moment in which the electrolyte has been emptied in the cell. At this time the current source is still turned off. When there are no changes in the PA amplitude, the current source is turned on.  As can be seen, an increase in the amplitude of the photoacoustic signal immediately appears. If the etching time increases, the photoacoustic signal becomes periodical (see Figure~\ref{Photoacustic1234}(b) to ~\ref{Photoacustic1234}(d)).  From this figure, it is possible to calculate the etching times. The thickness of the PS can be determined by using SEM images and by using the maximum and minimum conditions given by Eq.~(\ref{DELTA}) it is possible to determine the real part of the refractive index at this wavelength.

Taking into account the Eq.~\ref{LSource}, the change in the temperature of the sample is governed by the apparition of the porosity and changes in the thickness of the sample as shown in Eqs.~(\ref{Reflectance}) and~(\ref{DELTA}). From a physics point of view, the reflectance during the etching process is periodic (see Figure 3 (a)). Therefore, the changes in the amplitude of the photoacoustic signal are produced by the changes in the reflectance. Maximums in the reflectance correspond with a minimum in the photoacoustic signal, and minimums in the reflectance correspond with the maximum absorption of the radiation in the structure (PS/Si).

Here, it is imperative to clarify that the periodicity in the photoacoustic signal does not mean that the growing process is periodic. The PA signal is due to the light absorption and changes in the sample due to the modulation of the temperature.   At this point, the chemical reaction is considered constant during the process.

\begin{figure}
\includegraphics[scale=0.2]{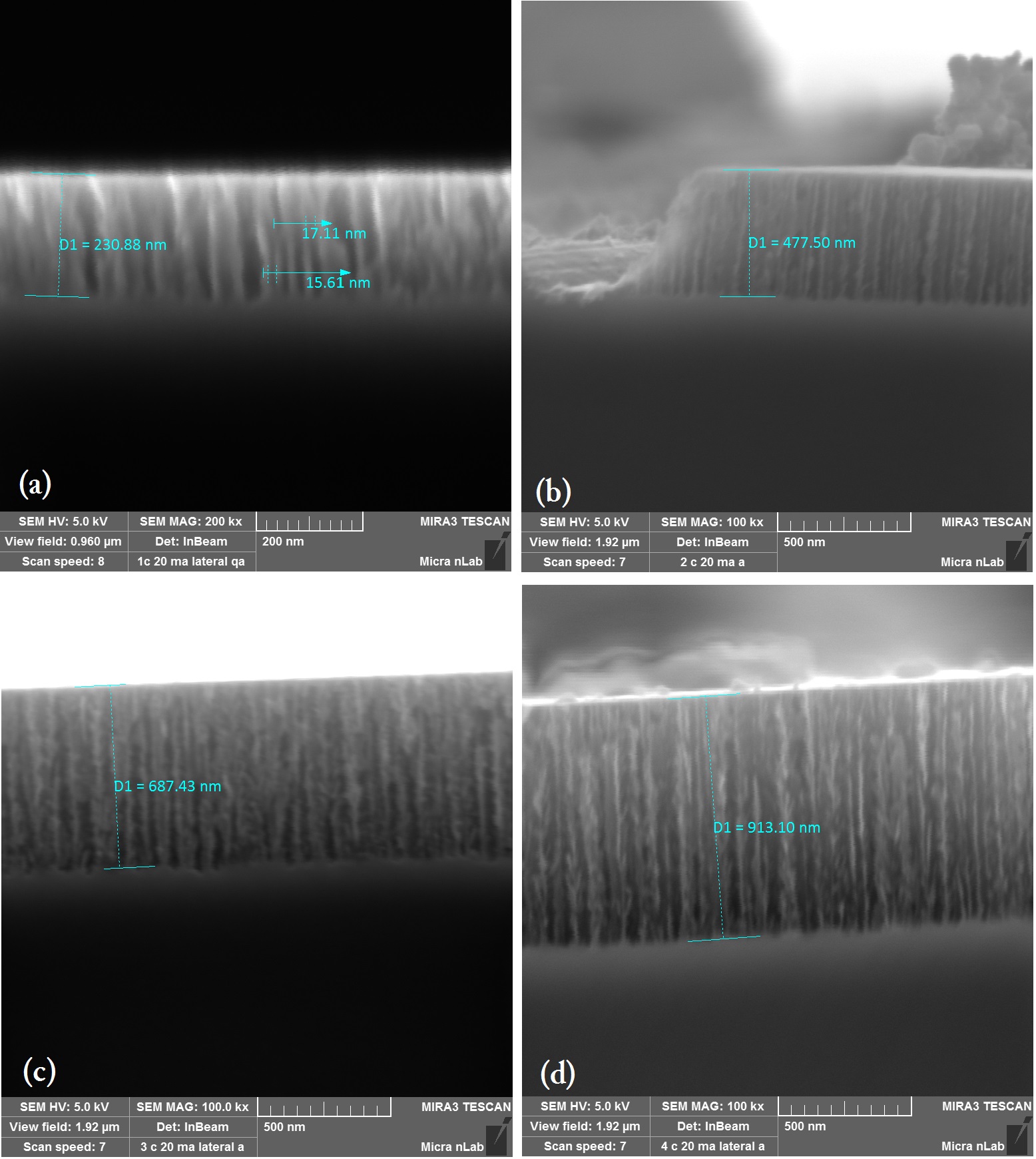}
\caption{ SEM images of cross sections of PS films growing for 1(a), 2(b), 3(c), and 4(d) photoacoustic cycles }
\label{SEM1234}
\end{figure}
\begin{figure}
\includegraphics[scale=0.3]{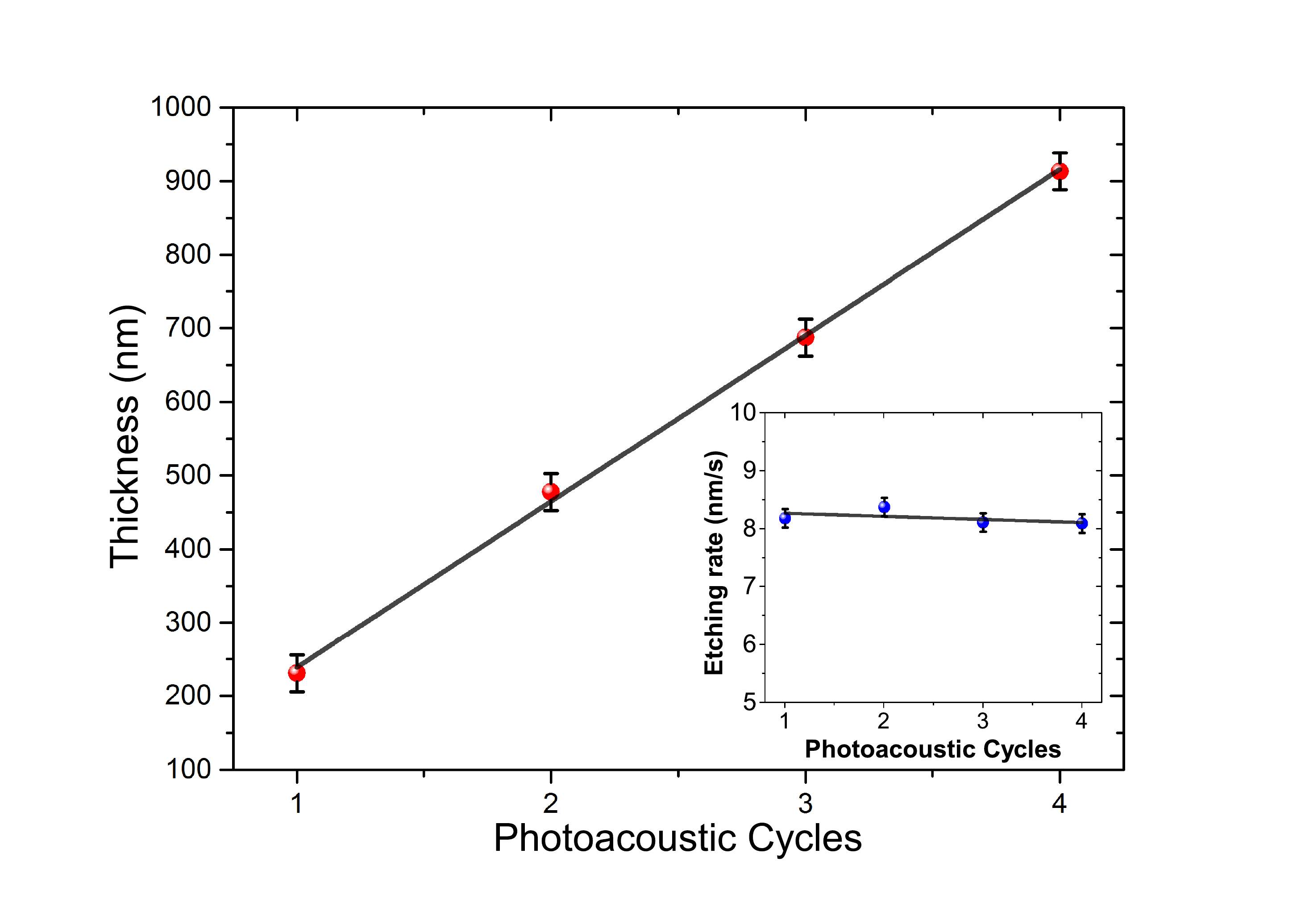}
\caption{Relationship between photoacoustic cycles and the thickness of the sample. The inset shows that the etching rate for these anodization conditions is  $8.3\pm0.1$ nm/s }
\label{ThickVsPA}
\end{figure}
To make a correlation between the photoacoustic signal for each period and the thickness of the samples, SEM images were obtained at the end of the 1st, 2nd, 3rd, and 4th photoacoustic periods. Figure~\ref{SEM1234} shows the cross sections of the PS films. As can be seen, the PS growing process produces nonregular Si columns in the range of 5  and 20 nm. The end of the pores are shown to be concave. From the data shown in Figure~\ref{Photoacustic1234}, is possible to determine the value of the period of the photoacoustic signal, and from Figure~\ref{SEM1234} it is possible to establish the average value of the sample thickness.

\begin{table*}
\caption{\label{tab:table1}Values of the real part of the refractive index  and the etching rate for the films of PS obtained during 1, 2 , 3, and 4 photoacoustic cycles. The anodization current was 20 mA and the ratio Ethanol/HF was 7:3 in volume.}
\begin{ruledtabular}
\begin{tabular}{ccccc}
 Photoacoustic cycles & Laser wavelength (nm) &Thickness (nm) &  Re$\left(\eta\left(\lambda\right)\right)$ &Etching rate nm/s $\pm0.1$\\ \hline
1&808&230.88\footnote{Direct determination by SEM, see Figure~\ref{SEM1234}. \label{SEMthick}}&1.75&8.1 \\
2&808&477.50\footref{SEMthick}&1.76    & 8.3\\
3&808&687.43\footref{SEMthick}&1.76    &8.1 \\
4&808&913.10\footref{SEMthick}&1.76    &8.0 \\
\end{tabular}
\end{ruledtabular}
\end{table*}
It is clear that a linear relationship exists between the thickness of the sample and the period of the photoacoustic signal (see Figure~\ref{ThickVsPA}); by the obtainment of the time for each one of the periods and thickness of the sample, the etching rate can be determined.  From the inset shown in the Figure~\ref{ThickVsPA}, the conclusion can be made that the etching rate is constant along the growing process, and that by using this methodology it is possible to obtain PS films with pre-established thickness values. Using the data obtained for the thickness of each sample, it was possible to obtain the value of the real part of the refractive index for the wavelength of the laser ($\lambda_0=808$  nm).  Table 1 shows the computed values obtained using the Eq.~(\ref{DELTA}). According to this table, it shows clearly that if the sample is grown under the same experimental conditions, the photoacoustic methodology shown for all cases has the same period for the amplitude of the signal. This is an important fact because this method allows the obtainment of samples with the same optical properties. Samples with the same photoacoustic history should be samples with similar physical properties.

\subsection{Photoacoustic signal wavelength dependence}
\begin{figure}
\includegraphics[scale=0.37]{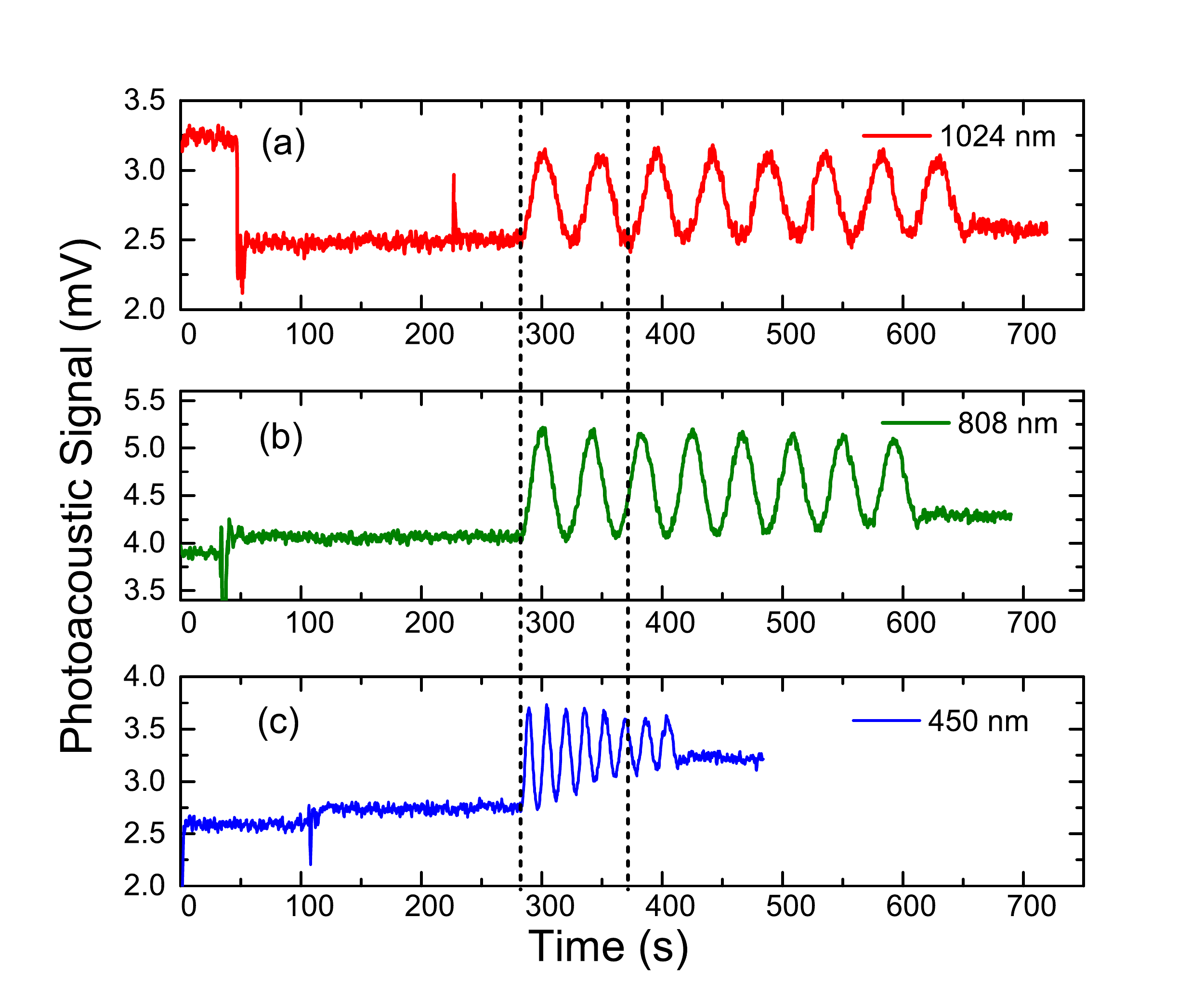}
\caption{Photoacoustic signal wavelength dependence. Three different samples growth under the same anodization condition but monitoring with lasers with different wavelengths.}
\label{PAWavelength}
\end{figure}

Another experiment was carried out to prove that the periodic behaviors of the reflectance are correlated to the photoacoustic period. Of course, the reflectance is dependent on the wavelength of the incident radiation. For this reason, three samples were grown using the same anodization conditions, but now using three different wavelengths: 1024, 808, and 450 nm. By a detailed analysis of the Eq.~(\ref{DELTA}), the period of the photoacoustic signal must increase if the incident wavelength increases. Figure~\ref{PAWavelength} (a) to (c) show the photoacoustic amplitude signal for three different wavelengths as a function of the time for eight photoacoustic cycles. It is clear that the value of the period in each case is different, and the experiment proved the correlation between the wavelength and the photoacoustic period proposed in the Eq. (\ref{DELTA}). This is a very important fact, because the most precise determination of the sample thickness can be achieved using a laser with a shorter wavelength.

\section {CONCLUSIONS}
Considering the above mentioned results, it can be concluded that the photoacoustic signal in the case PS formation  is governed by the changes in the reflectance of the system (PS/Si). An important point here is that it is necessary to explore the contribution of the electrochemical reaction. A detailed inspection of Figure~\ref{ThickVsPA} in which there is clear a linear relationship between PS thickness and photoacoustic signal, suggests that  the chemical reaction should be constant along the PS formation nonly if the electrolyte does not suffer significant changes in concentration.

The photoacoustic methodology proposed in this study permits the obtainment of PS samples with the same characteristics assuming no changes occurred in the impurities across the sample. This means that it is possible to determine the etching rate, and using at least one SEM image, it is possible to obtain the effective refractive index of the sample. Comparing the reflective index with any model of an effective medium such as (Looyenga\cite{Looyenga}, Bruggeman\cite{Bruggeman}, Maxwell-Garnet\cite{Maxwell}), this methodology can also determine the porosity of the sample. The method permits monitoring of the etching process using different laser wavelengths. This fact can be used to calculate the refractive index for these wavelengths and constructs a real curve of effective medium.  Another important issue is the direct relationship between the laser wavelength and the period of the amplitude of the photoacoustic signal. In the case of the development of optical devices such as Bragg reflectors or Fabry-Perot cavities, in which the thickness and porosity play an important role, this methodology can be used.

\begin{acknowledgments}
C. F. Ramirez-Gutierrez and J. D. Casta\~no-Yepes want to thank Consejo Nacional de Ciencia y Tecnolog\'ia M\'exico, for the financial support of their Ph.D. studies. This work was supported by PAPIIT IN115113 UNAM-M\'exico.  All the authors want to thank Micra Nanotecnolog\'ia SA de CV for the support in the electron microscopy performance and  M.Sc. Julio Cesar Franco Correa for the  support  in the photoacoustic experiments.
\end{acknowledgments}


\begin{thebibliography}{X}
\bibitem{1} S. E. Foss, P. Y. Y. Kan, and T. G. Finstad, J. Appl. Phys. \textbf{97}, 114909 (2005).
\bibitem{2} L. T. Canham, T. I. Cox, A. Loni, and A. J. Simons, Appl. Surf. Sci. \textbf{102}, 436 (1996).  
\bibitem{3} P. A. Snow, E. K. Squire, P. St. J. Russell,  and L. T. Canham, J. Appl. Phys. \textbf{86}, 1781 (1999).
\bibitem{4} J. Charrier and M. Dribek, J. Appl. Phys. \textbf{107}, 044905 (2010).
\bibitem{5} Y. Y. Li, F. Cunin, J. R. Link, T. Gao, R . E. Betts, S. H. Reiver, V. Chin, S. N. Bhatia, and M. J. Sailor,  Science  \textbf{299}, 5615, 2045-2047  (2003).
\bibitem{Gutierrez} A. Guti\'errez, J. Giraldo, R. Vel\'azquez-Hern\'andez, M. L. Mendoza-L\'opez, D. G. Espinosa-Arbel\'aez, A. del Real, and M. E. Rodriguez-Garc\'ia, Rev. Sci. Instrum. \textbf{81}, 013901 (2010).
\bibitem{Espinosa} D. G. Espinosa-Arbel\'aes, R. V. Vel\'asquez-Hern\'andez, J. Petricioli-Carranco, R. Quintero-Torres, and M. E. Rodr\'iguez-Garc\'ia, Phys. Status Solidi C. \textbf{8}, 6, 1856-1859 (2011).
\bibitem{Kan} P. Y. Y. Kan, S. E. Foss, and T. G. Finstad, Phys. Status Solidi A. \textbf{202}, 8, 1533-1538 (2005).
\bibitem{RGPhoto} A. Rosencwaig and A. Gersho, J. Appl. Phys. \textbf{47}, 64 (1976).
\bibitem{Mitsas} C. L.  Mitsas and D. I. Siapkas, Appl. Opt. \textbf{34}, 10, 1678-1683 (1995).
\bibitem{RCA1} M. Itano, F. W. Kern, M. Miyashita, and T Ohmi,  IEEE Trans. Semicond. Manuf. \textbf{6} (3), 258-267 (1993).
\bibitem{RCA2} W. Kern, J. Electrochem. Soc. \textbf{137} (6), 1887-1892 (1990).
\bibitem{Looyenga} H. Looyenga,  Physica \textbf{31} (3), 401-406 (1965).
\bibitem{Bruggeman} D. A. G. Bruggeman, Ann. Phys. \textbf{24}, 636 (1935).
\bibitem{Maxwell} G. A. Niklasson, C. G. Granqvist, and O. Hunderi,  Appl. Opt. \textbf{20}, 1,26-30 (1981).
\end{thebibliography}
\end{document}